\documentclass[12pt,twoside,a4paper]{article} 
\usepackage{natbib}
\usepackage{bled2e}
\usepackage{amsmath}
\usepackage{bm}
\usepackage[justification=centering,margin=10pt,font=small,labelfont=bf]{caption}

\usepackage{graphicx}
\usepackage{array}

\usepackage{times}

\makeatletter
\@addtoreset{equation}{section}
\makeatother
\title{A note on the normality assumption for Bayesian models of constraint in behavioral individual differences}
\author{
  Thomas J. Faulkenberry\thanks{~~Department of Psychological Sciences, Tarleton State University; faulkenberry@tarleton.edu}
}

\date{ }
\begin{document}

\maketitle

\markboth{Thomas J. Faulkenberry}
         {A note on the normality assumption \ldots}

\begin{abstract}
\setcounter{footnote}{1}
To investigate the structure of individual differences in performance on behavioral tasks, \citet{haaf2017} developed a class of hierarchical Bayesian mixed models with varying levels of constraint on the individual effects. The models are then compared via Bayes factors, telling us which model best predicts the observed data. One common criticism of their method is that the observed data are assumed to be drawn from a normal distribution. However, for most cognitive tasks, the primary measure of performance is a response time, the distribution of which is well known not to be normal. In this paper, I investigate the assumption of normality for two datasets in numerical cognition. Specifically, I show that using a shifted lognormal model for the response times does not change the overall pattern of inference. Further, since the model-estimated effects are now on a logarithmic scale, the interpretation of the modeling becomes more difficult, particularly because the estimated effect is now multiplicative rather than additive. As a result, I recommend that even though response times are not normally distributed in general, the simplification afforded by the \citet{haaf2017} approach provides a pragmatic approach to modeling individual differences in behavioral tasks.
\end{abstract}

%\newpage

\section{Introduction}

In the behavioral sciences, a common target of investigation is individual performance on behavioral tasks, and particularly whether this individual performance can predict other measureable outcomes. As an illustrative example, consider a simple number comparison task where subjects are asked to choose the physically larger of two number digits presented in different physical sizes on a screen (e.g., a large numeral 2 displayed alongside a small numeral 8). Performance on this task routinely exhibits a \emph{size congruity effect} \citep{henik1982}, where people are slower (on average) to choose the larger when the numbers are presented in a physical size configuration that is incongruent with their relative numerical magnitude. Importantly, researchers often use individual performance on this task to predict other meaningful behavioral outcomes, especially those related to mathematics anxiety and ability. For example, \citet{rubinsten2005} found that people with developmental dyscalculia exhibited a smaller size congruity effect compared to a typical-functioning control group, which they interpreted as evidence for a lack of automatic number activation in the dyscalculia group.\\

Given that individual performance on behavioral tasks is used in this metric sense to predict other tangible outcomes, a natural question concerns whether this metric scale has any constraint. Specifically, \citet{haaf2017} proposed that a method to ascertain whether people differ in performance solely in a {\it quantitative} fashion (i.e., everybody exhibits the effect in the same direction, but differ in the size of the effect), or whether there are also {\it qualitative} individual differences, where some people exhibit a positive effect, but others exhibit the effect in the opposite direction. Such questions are becoming important in the psychological and behavioral sciences, particularly in terms of providing much-needed constraint on the plethora of observed ``effects'' and helping to guide more targeted theoretical development about human behavior \citep{rouder2021}.  \\

The purpose of this paper is to examine one of the fundamental statistical assumptions of the \citet{haaf2017} method for modeling constraint on behavioral individual differences. In brief, the method relies on assuming that the resultant behavioral measures (e.g., response times) are drawn from a normal distribution whose mean is represented as a linear combination of a variable intercept and slope (effect). In turn, each of these parameters is further drawn from normal distributions centered at 0 and scaled according to overall variability. Different models of individual difference structure are instantiated by placing varying levels of constraint on the slope/effect parameter. Critical to the \citet{haaf2017}  method is a Bayesian comparison of these models, which uses a combination of the well-known analysis of variance approach developed by \citet{rouder2012} and the encompassing prior approach \citep{faulkenberry2019encompassing}. The approach has been used successfully to investigate the structure of individual differences in many behavioral phenomena, including Stroop and Simon effects \citep{haaf2018}, the truth effect \citep{schnuerch2020}, numerical distance effects \citep{vogel2021}, and the numerical size congruity effect \citep{faulkenberryBowman2020}.\\

One criticism of the \citet{haaf2017} method is the assumption that raw performance measures are drawn from a normal distribution. This criticism is particularly salient when the primary measure is response time, as response times are well known to exhibit a distinct positive skew. While there are many methods for modeling response times using skewed distributions (i.e., ex-Gaussian, inverse Gaussian / Wald, etc.), the implementation of such distributions into the \citet{haaf2017} framework is quite difficult. Compared to a normal distribution, these distributions involve multiple parameters, and so it is not clear on which parameter the congruity effect should be applied. Further, the \citet{haaf2017} method is specifically built on an established computational framework built into the BayesFactor package in R, so deviating from this method would require the user to develop many new techniques from first principles. As an alternative, one simple approach that might prove attractive is to assume that the observed response times follow a (shifted) lognormal distribution; then, the analyst may simply transform the observed response times by first shifting by a fixed amount (e.g., 200 milliseconds is a common recommendation) and then taking the (natural) logarithm. The resulting distribution of (log) response times is then approximately normal and may be ``fed into'' the \citet{haaf2017} method with little difficulty.\\

The purpose of this brief paper is to investigate this approach and to argue two points:

\begin{enumerate}
\item the inferences obtained from the shifted lognormal model are practically equivalent as those obtained from the original model with the normality assumption; and
\item interpreting the estimated model parameters from the shifted lognormal model is nontrivial and potentially inappropriate in the context of these behavioral tasks.
\end{enumerate}

\section{Bayesian model implementation}

First, I will describe the Bayesian mixed model approach developed by \citet{haaf2017}, particularly as applied to behavioral tasks where the primary observed data are response times. Before going into the details, I will reiterate that the main aim of this approach is to build a (hypothetical) generative process for each observed response time in a behavioral task. That is, there is no aggregation of trials at the individual or group level that needs to occur.\\

Each observed response time is assumed to be the sum of four components: (1) a grand mean \(\mu\); (2) a subject-specific adjustment \(\alpha\) to the grand mean (i.e., so that \(\mu+\alpha\) gives a ``random'' intercept for each subject); (3) a subject-specific effect term \(\delta\); and (4) a noise term \(\varepsilon\). The hierarchical model is then built by assuming each of these components is drawn from some to-be-defined probability distribution. Of particular interest is the distribution that generates each subject's effect term -- this distribution is the one on which we build our competing models of individual difference structure.\\

We let \(Y_{ijk}\) denote the response time for the \(k^{\text{th}}\) replicate of the \(i^{\text{th}}\) subject in the \(j^{\text{th}}\) experimental condition (usually two conditions, so that \(j=1,2\)). As described, our random effects linear model on the vector of response times \(Y_{ijk}\) looks like:
\[
Y_{ijk} \sim \mathcal{N}(\mu + \alpha_i + x_j\cdot \delta_i, \sigma^2).
\]
Here, \(\mu\) denotes the grand mean intercept and \(\alpha_i\) represents the specific intercept adjustment for subject \(i\). The term \(x_j\) is a binary variable which codes the experimental condition for each trial. For example, suppose we are interested in modeling a congruity effect, where response times on incongruent trials generally increase compared to those of congruent trials. In this case, for congruent trials (condition \(j=1\)), we would set \(x_1=0\), and for incongruent trials (condition \(j=2\)), we would set \(x_2=1\). Under such a specification, \(\delta_i\) then represents the (random) congruity effect for subject \(i\). Finally, \(\sigma^2\) represents the latent sampling variance of the observed response times.

The next step is to propose a structure for the parent distribution of random effects \(\delta_i\) (i.e., the distribution from which each subject's size-congruity effect \(\delta_i\) is randomly drawn). We define four possible populations for these \(\delta_i\), each of which mathematically specifies one of four possible theoretical positions about the distribution of effects.

\subsection{The unconstrained model}

The unconstrained model, denoted \(\mathcal{M}_u\), allows the effects \(\delta_i\) to vary both in type/quality (i.e., positive or negative) as well as magnitude. As such, with \(\mathcal{M}_u\) we place no constraint on the individual effects \(\delta_i\). We define this model as
\[
\mathcal{M}_u: \delta_i \sim \mathcal{N}(\nu, \eta^2),
\]
where \(\nu\) and \(\eta^2\) represent the mean and variance, respectively, of the distribution of individual effects \(\delta_i\).

\subsection{The positive-effects model}

The positive-effects model, denoted \(\mathcal{M}_+\), hypothesizes that effects \(\delta_i\) only vary in quantity (i.e., they are always positive, but possibly differ in magnitude between subjects). \(\mathcal{M}_+\) is a constrained model in the sense that it specifies the assumption that all individual effects \(\delta_i\) are positive. That is,
\[
\mathcal{M}_+:\delta_i \sim \mathcal{N}_+(\nu,\eta^2),
\]
where \(\mathcal{N}_+\) denotes a truncated normal distribution with lower bound 0.

\subsection{The common-effect and null models}

Whereas the unconstrained and positive-effects models are usually the primary players in studies on individual structure, the common-effect and null models are defined to provide a critical check of experimental design. The common-effect model places even more constraint on the distribution of effects by assuming that each individual has the \emph{same} effect. That is,
\[
\mathcal{M}_1:\delta_i = \nu,
\]
Such a model serves to probe the following question: if the common-effect model was the best predictor of the observed data, one would be forced to question the efficiency of the experimental design as a test to elicit individual differences in the effect. As one might expect, the null model is the most constrained of the four, as it specifies that each subject's size-congruity effect is zero:
\[
\mathcal{M}_0:\delta_i = 0.
\]
It is used for a similar reason: if the null model was the best predictor of the observed data, then one must question the efficiency of the experimental design to elicit effects of any sort.

\subsection{Prior specifications}

Generally, most applications of the \citet{haaf2017} method follow similar ``default'' prior specifications. The critical parameters I'll describe here are \(\delta_i\), \(\nu\), and \(\eta^2\). The default procedure is to use the \(g\)-prior approach \citep{rouder2012,zellner1986}, which re-expresses these parameters as a standardized effect size. To see how this works, consider the collection of individual effect parameters \(\delta_i\). We define \(g_{\delta} = \eta^2/\sigma^2\), yielding a hyperparameter that casts the variability of \(\delta_i\) in terms of the ratio of true variability \(\eta^2\) to sampling variability \(\sigma^2\). With this we can re-write our unconstrained model as
\[
\mathcal{M}_u:\delta_i \sim \mathcal{N}(\nu, g_{\delta}\sigma^2).
\]
Similarly, we may scale the mean size-congruity effect \(\nu\) in terms of sampling variability and get a new hyperparameter \(g_{\nu}\). Continuing up the hierarchy, these new (hyper)parameters need priors as well. The default specification (Zellner, 1986) is to use scaled \(\text{inverse-}\chi^2\) distributions with one degree of freedom and scale \(r^2\).\\

To be clear, the \(g\)-prior setup is quite clever, as it completely describes these critical parameters in terms of sampling variability \(\sigma^2\). By doing this, we convert the problem of specifying priors on \(\delta_i\), \(\nu\), and \(\eta^2\) into one where we simply need to specify the expected variability of our effects relative to the expected overall variability of the observed response times. Like \citet{haaf2017}, I will use \(\sigma=300\) milliseconds as a prior expectation for the variability of observed response times.\\

Now we can actually finish setting our priors. First we consider \(g_{\nu}\), the \(g\)-prior on the mean size-congruity effect. With the \(g\)-prior setup, we assume that \(\nu \sim \mathcal{N}(0, g_{\nu}\sigma^2)\), where \(g_{\nu} \sim \text{Scale-inv-}\chi^2(r_{\nu}^2)\). The scale parameter \(r_{\nu}\) should reflect our prior belief about the relative magnitude of our expected effects. For the types of effects we often see in numerical cognition (and certainly the types of tasks we will describe in this paper), I usually expect such effects to be, on average, around 50 milliseconds, or 1/6 of the expected overall trial-by-trial variability (\(\sigma = 300\) milliseconds). Thus, we set \(r_{\nu}=1/6\).\\

Second, we consider \(g_{\delta}\), which describes the variability of individual effects around the mean effect. With the \(g\)-prior setup, we assume that \(g_{\delta} \sim \text{Scale-inv-}\chi^2(r_{\delta}^2)\). Like \citet{haaf2017}, we set \(r_{\delta}=1/10\), which would indicate that the expected variability of the effect across individuals should be about 1/10 of \(\sigma=300\) milliseconds, or around 30 milliseconds.

\subsection{Model comparison}

Since our goal is to capture the latent structure of individual differences in the effects we observe in our behavioral task, our problem is first and foremost one of {\it model comparison}. That is, we ask which of the four competing models defined above is the most adequate as a predictor of our observed data? To answer this question, we use Bayes factors \citep{jeffreys1961, kass1995}, which index the relative predictive adequacy of two models by comparing the marginal likelihood of observed data under one model to another \citep{faulkenberry2020,faulkenberry2022}. For example, a Bayes factor of 10 indicates that the observed data are 10 times more likely under one model compared to another. Techniques for computing Bayes factors among three of the four models above (\(\mathcal{M}_u\), \(\mathcal{M}_1\), \(\mathcal{M}_0\)) were previously developed by \citet{rouder2012} and are implemented in the BayesFactor \citep{BayesFactor} package in R \citep{R}. The Bayes factor between the constrained positive effects model \(\mathcal{M}_+\) and the unconstrained model \(\mathcal{M}_u\) is computed by the \emph{encompassing prior} method \citep{klugkist2005,faulkenberry2019encompassing}, which is based on counting the number of posterior samples of \(\mathcal{M}_u\) which obey the constraint placed by \(\mathcal{M}_+\), then comparing this to the number of prior samples which obey the same constraint.

\section{Case studies}\label{case-studies}

My goal in this paper is to compare the inferences from the default \citet{haaf2017} method, which assumes that the observed response times are drawn from a \emph{normal distribution}, to a modified approach where the observed response times are drawn from a \emph{lognormal distribution}. To do this, I will perform two case studies where I analyze two datasets that have already appeared in the literature. In case study 1, I will model the latent structure of individual differences in the \emph{size congruity effect} \citep{henik1982}, a classic phenomenon in numerical cognition in which people are slower to choose the larger of two presented numbers when the numbers are presented in a physical size that is incongruent with their relative numerical magnitude (e.g., a large numeral 2 displayed alongside a small numeral 8). The data for case study 1 (19,499 response times from \(N=\) 53 subjects) were originally reported in \citet{faulkenberryBowman2020}. In case study 2, I will model the latent structure of individual differences in the \emph{unit decade compatibility effect}, another classic phenomenon in numerical cognition \citep{nuerk2001}. The data for case study 2 (11,600 response times from \(N=\) 53 subjects) are unpublished but available as part of a collaborative pregistration project by \citet{cipora2021}.

\subsection{Case study 1 -- size congruity effect}

\begin{figure}[htbp]
\centering
\includegraphics[width=\textwidth]{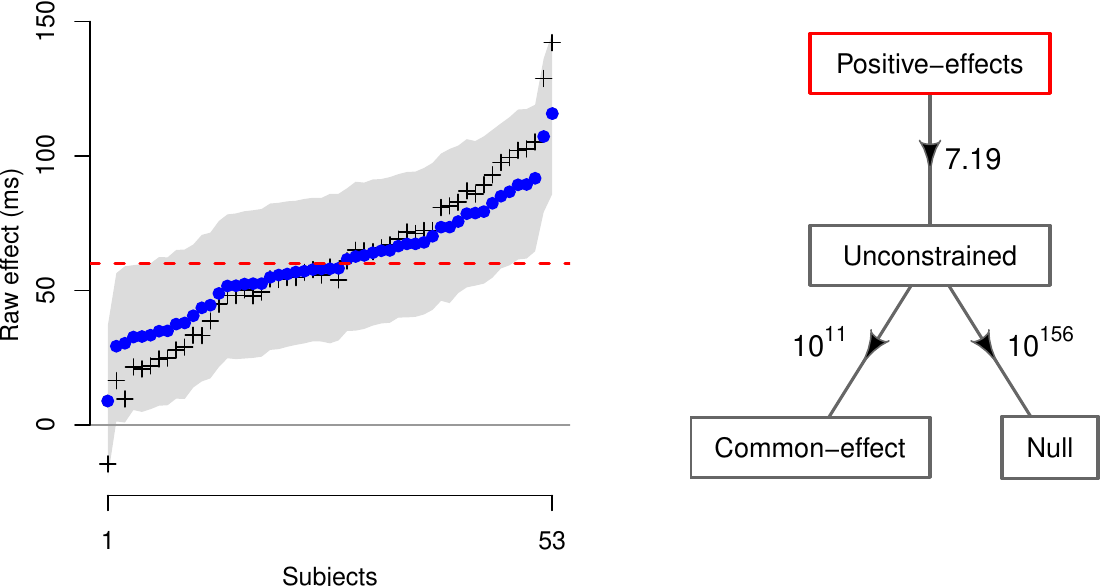}
\caption{\label{fig:plot1}Individual effect estimates (left column) and Bayes factor model comparisons (right column) for Case Study 1 under a normal distribution assumption. Posterior means and 95\% credible intervals for \(\delta_i\) are represented by blue dots and a gray band, respectively. The + symbols represent the observed size-congruity effect for each subject. The red dashed-line represents the estimated mean size-congruity effect \(\nu\). For the model comparisons, the red box denotes the winning model, and Bayes factors are displayed beside each arrow.}
\end{figure}

The first analysis I will describe is the default \citet{haaf2017} method, which places a normal distribution on the observed response times. The individual effect estimates from the unconstrained model are displayed in the left column of Figure \ref{fig:plot1}. We can see that the observed effects for each subject (denoted by black crosses) span from -14.59 ms to 142.10 ms. In this context, we compute observed effects by subtracting each subject's mean response time for congruent trials from the mean response time for incongruent trials. With the exception of one subject, the observed size-congruity effects are all constrained to be positive. Estimates from the hierarchical Bayesian model are displayed as blue dots with shaded 95\% credible intervals. These estimates are computed as means of the posterior samples for each \(\delta_i\), and the 95\% credible intervals are computed as the central 95\% of the posterior samples (i.e., ranging between the 2.5\% and 97.5\% quantiles of the samples). The red dashed line represents an (posterior) estimated mean effect of \(\nu=\) 60 ms.\\

As is usually seen with this type of modeling (and hierarchical modeling in general), we observe a fair amount of \emph{shrinkage} in our estimates. Notice that the estimated effects (the blue dots) extend over a smaller range (8.84 ms to 115.73 ms) than the observed effects (the black crosses; -14.59 ms to 142.10 ms). This shrinkage reflects how the hierarchical model accounts for sampling variability at all levels.\\

The right column of Figure \ref{fig:plot1} shows the Bayes factor model comparisons. As we can see, the observed data were 7.19 times more likely under the positive-effects model \(\mathcal{M}_+\) than under the unconstrained model \(\mathcal{M}_u\). If we assume 1-to-1 prior odds for \(\mathcal{M}_+\) and \(\mathcal{M}_u\), this means that our posterior odds in favor of \(\mathcal{M}_+\) have increased to 7.19-to-1, which is equivalent to a posterior probability of \(\text{Pr}(\mathcal{M}_+ \mid \text{data})=\) 0.88. These models were overhelmingly preferred over the common-effect model \(\mathcal{M}_1\) and the null model \(\mathcal{M}_0\), as \(\mathcal{M}_u\) was more likely to have predicted the observed data by factors of \(10^{11}\)-to-1 and \(10^{156}\)-to-1, respectively.\\

\begin{figure}[htbp]
\centering
\includegraphics[width=\textwidth]{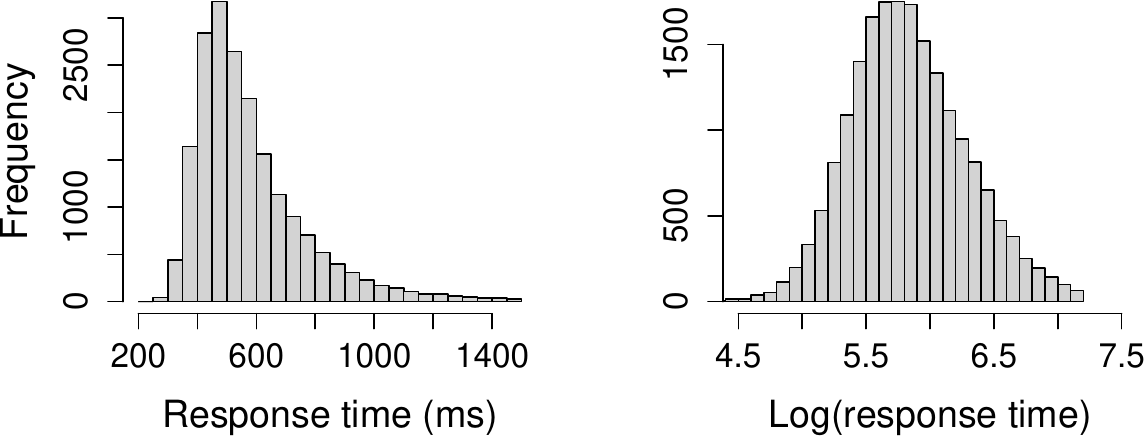}
\caption{\label{fig:plotTransform1}Distributions of observed response times in the size congruity task (Case study 1). The left panel displays the original observed response times, whereas the right panel displays the log-transformed response times.}
\end{figure}

Next, we perform the same procedure while assuming a shifted lognormal distribution on the observed response times. To do this, we transform the observed response times by first subtracting a constant amount from each response time (here, I chose a shift of 200 milliseconds), then taking the (natural) logarithm of the result. As we can see in Figure \ref{fig:plotTransform1}, the transformed distribution appears approximately normal, indicating that the lognormal model is appropriate in this case study as well. The resulting transformed data can be directly modeled as above, the results of which I will now describe.\\

The overall similarity of these results with the first analysis is striking. We see very similar patterns of observed effects, estimated effects, and shrinkage. For the log transformed data, we see a posterior estimated common effect (red dashed line) \(\nu=\) 0.15. If we back-transform this back to the original response time scale, we get an estimated common effect of 1.16. Because the data are on a logarithmic scale, this effect is multiplicative, so an estimated effect of 1.16 is a 16\% increase in response times. For these data, this is roughly equivalent to a response time increase of 86 ms.\\

\begin{figure}[htbp]
\centering
\includegraphics[width=\textwidth]{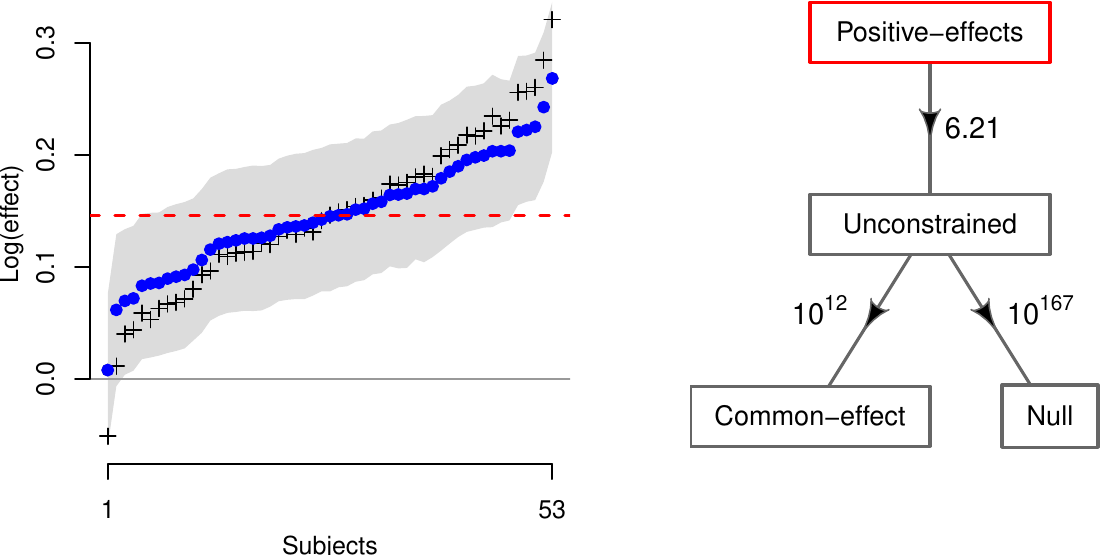}
\caption{\label{fig:plot1log}Individual effect estimates (left column) and Bayes factor model comparisons (right column) for Case Study 1 under a \emph{lognormal} distribution assumption. Posterior means and 95\% credible intervals for \(\delta_i\) are represented by blue dots and a gray band, respectively. The + symbols represent the observed size-congruity effect for each subject. The red dashed-line represents the estimated mean size-congruity effect \(\nu\). For the model comparisons, the red box denotes the winning model, and Bayes factors are displayed beside each arrow.}
\end{figure}

The similarity persists with the Bayes factor comparisons. In the right column of Figure \ref{fig:plot1log} we can see the observed data were 6.21 times more likely under the positive-effects model \(\mathcal{M}_+\) than under the unconstrained model \(\mathcal{M}_u\). Further, these models were again overhelmingly preferred over the common-effect model \(\mathcal{M}_1\) and the null model \(\mathcal{M}_0\).\\

In all, it seems that with the exception of the raw effect estimate, the inferences we obtain from using a shifted lognormal model on observed response times is very similar to that when we use the default normal specifications recommended by \citet{haaf2017}. In both cases, the positive effects model is preferred over the unconstrained model.

\subsection{Case study 2 - unit decade compatibility effect}

\begin{figure}[htbp]
\centering
\includegraphics[width=\textwidth]{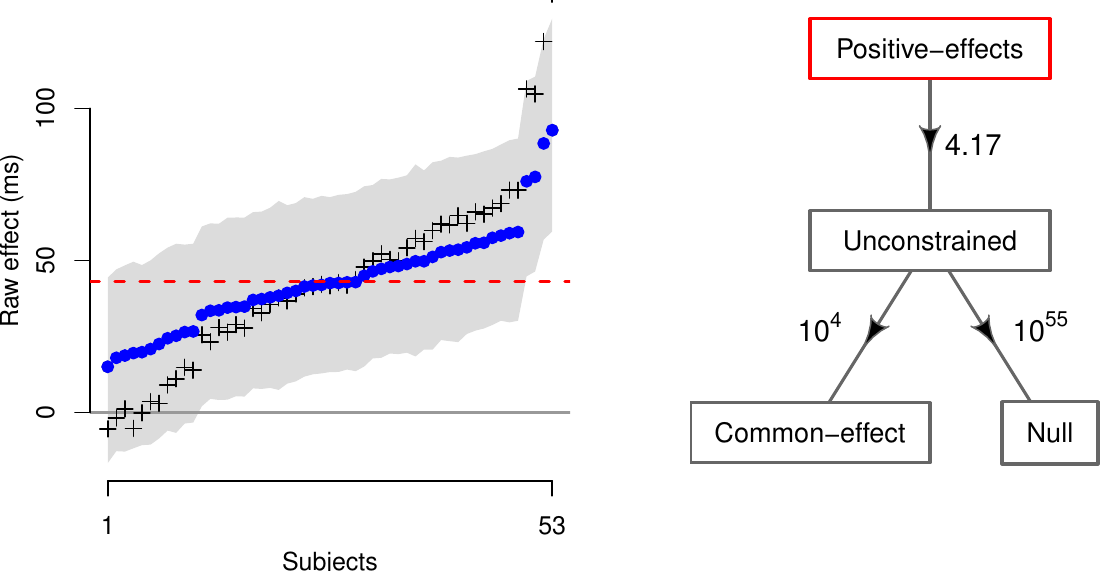}
\caption{\label{fig:plot2}Individual effect estimates (left column) and Bayes factor model comparisons (right column) for Case Study 2 under a normal distribution assumption. Posterior means and 95\% credible intervals for \(\delta_i\) are represented by blue dots and a gray band, respectively. The + symbols represent the observed size-congruity effect for each subject. The red dashed-line represents the estimated mean size-congruity effect \(\nu\). For the model comparisons, the red box denotes the winning model, and Bayes factors are displayed beside each arrow.}
\end{figure}

As above, I will first report the results of modeling using the default \citet{haaf2017} method with a normal distribution on the observed response times. The individual effect estimates from the unconstrained model are displayed in the left column of Figure \ref{fig:plot2}. The observed effects for each subject (denoted by black crosses) span from -5.37 ms to 137.52 ms. Similar to Case Study 1, the observed effects were mostly positive. Estimates from the hierarchical Bayesian model are displayed as blue dots with shaded 95\% credible interval. The red dashed line represents an (posterior) estimated mean effect of \(\nu=\) 43 ms. Note that we again observe shrinkage in our estimates, as the estimated effects extend from 15.04 ms to 92.80 ms), a smaller range that that of the observed estimates.\\

The right column of Figure \ref{fig:plot2} shows the Bayes factor model comparisons. In this case, the observed data were 4.17 times more likely under the positive-effects model \(\mathcal{M}_+\) than under the unconstrained model \(\mathcal{M}_u\). If we assume 1-to-1 prior odds for \(\mathcal{M}_+\) and \(\mathcal{M}_u\), this means that our posterior odds in favor of \(\mathcal{M}_+\) have increased to 4.17-to-1, which is equivalent to a posterior probability of \(\text{Pr}(\mathcal{M}_+ \mid \text{data})=\) 0.81. As in Case Study 1, these models were strongly preferred over the common-effect model \(\mathcal{M}_1\) and the null model \(\mathcal{M}_0\).\\

\begin{figure}[htbp]
\centering
\includegraphics[width=\textwidth]{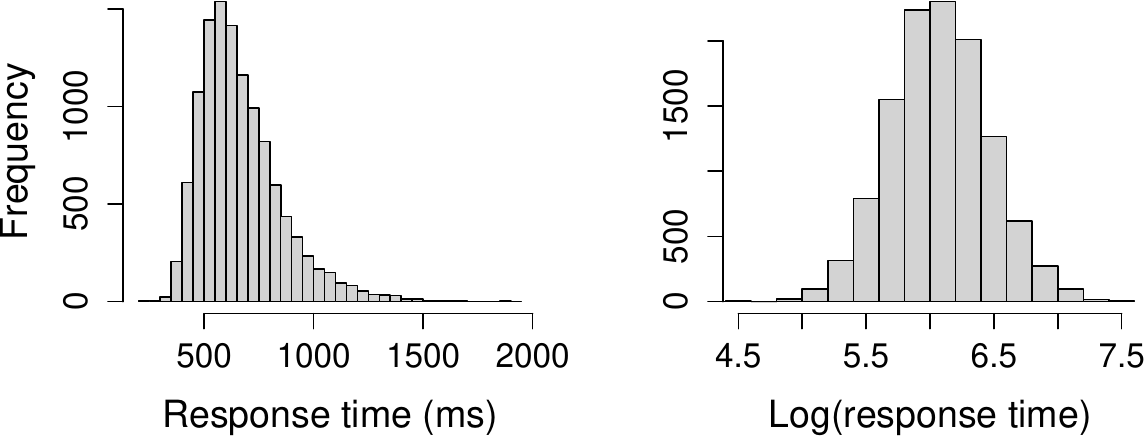}
\caption{\label{fig:plotTransform2}Distributions of observed response times in the numerical comparison task (Case study 1). The left panel displays the original observed response times, whereas the right panel displays the log-transformed response times.}
\end{figure}

Next, we run the analysis again, but this time assuming a shifted lognormal distribution on the observed response times. As before, we transform the observed response times by first subtracting a constant amount from each response time (here, I chose a shift of 200 milliseconds), then taking the (natural) logarithm of the result. As we can see in Figure \ref{fig:plotTransform2}, the transformed distribution appears approximately normal, indicating that the lognormal model is appropriate here.\\

As with Case Study 1, we see very similar patterns of observed effects, estimated effects, and shrinkage in Figure \ref{fig:plot2}. For the log transformed data, we see a posterior estimated common effect (red dashed line) \(\nu=\) 0.08. On the original response time scale, this is equivalent to an estimated (multiplicative) common effect of 1.09, a 9\% increase in response times. For these data, this is roughly equivalent to a response time increase of 57 ms.\\

\begin{figure}[htbp]
\centering
\includegraphics[width=\textwidth]{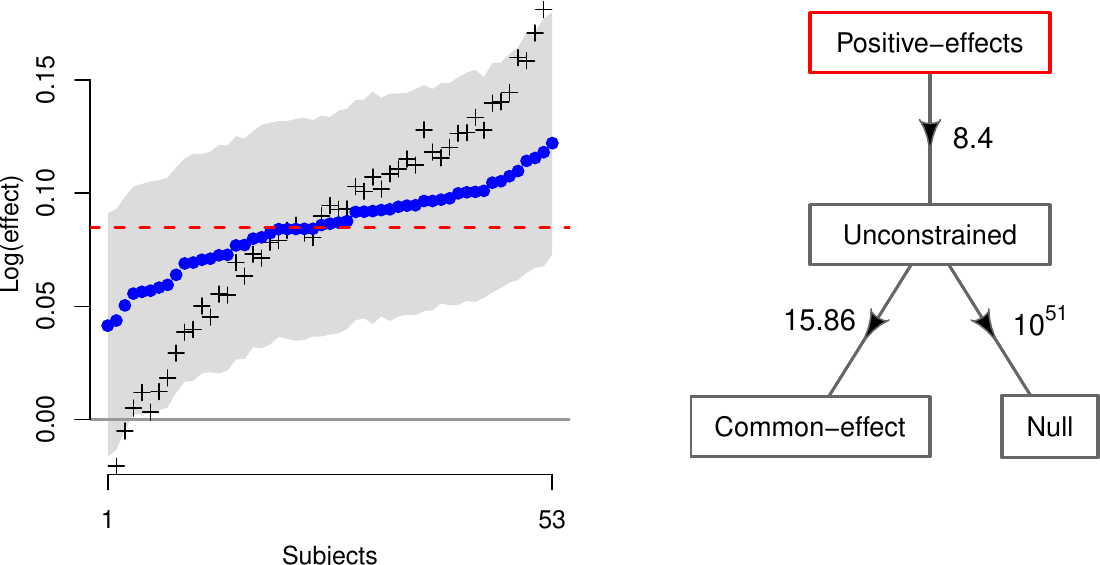}
\caption{\label{fig:plot2log}Individual effect estimates (left column) and Bayes factor model comparisons (right column) for Case Study 2 under a \emph{lognormal} distribution assumption. Posterior means and 95\% credible intervals for \(\delta_i\) are represented by blue dots and a gray band, respectively. The + symbols represent the observed size-congruity effect for each subject. The red dashed-line represents the estimated mean size-congruity effect \(\nu\). For the model comparisons, the red box denotes the winning model, and Bayes factors are displayed beside each arrow.}
\end{figure}

The Bayes factor comparisons also present the same message. In the right column of Figure \ref{fig:plot2log} we can see the observed data were 8.40 times more likely under the positive-effects model \(\mathcal{M}_+\) than under the unconstrained model \(\mathcal{M}_u\). Further, these models were again preferred over the common-effect model \(\mathcal{M}_1\) and the null model \(\mathcal{M}_0\). Again, the inference from using a shifted lognormal model on observed response times is very similar to that when we use the default normal specifications recommended by Haaf \& Rouder (2017). In both cases, the positive effects model is preferred over the unconstrained model.

\section{Simulation study}

As we can see from the previous two sections, both case studies lead to a common conclusion. Even though the observed response times exhibit positive skew, the inference we obtain from applying the default \citet{haaf2017} method (which assumes a normal distribution on response times) is practically the same as when we apply a shifted lognormal model on response times. To extend support for this tentative conclusion, I performed a simulation study to benchmark and compare the long-term performance of both methods against data which are assumed to be generated from either the positive-effects model or the unconstrained model. In this section, I describe the simulation study and report its results. The simulation was performed in R, and the simulation script can be viewed at https://bit.ly/3D9QGlZ.

For each simulation run, the data were assumed to be generated from a hierarchical shifted-Wald distribution \citep[e.g.,][]{anders2016,faulkenberry2018wald}. In general, a shifted-Wald distribution represents the collection of stopping times for a continuous accumulator with drift toward a fixed response boundary. The distribution exhibits a positive skew characteristic of response time distributions, which positions it as a good model for use here. The shifted-Wald can be completely described by three parameters: drift rate $\gamma$, which represents the rate at which information is accumulated during stimulus presentation; response threshold $\alpha$, which represents the amount of information that must be accumulated before a decision can be initiated; and shift $\theta$, which represents the remaining portion of the response time on a trial which is not accounted for by the accumulation process (i.e., perceptual encoding, motor preparation, etc.).

To explain how the data were simulated, consider the context of a typical repeated-measures experiment with an equal number of congruent and incongruent trials. Let us assume that $N$ subjects each produce $k$ trials in the congruent condition. The $k$ ``observed'' RTs for each subject $i=1,\dots,N$ are assumed to be randomly drawn from a subject-specific shifted-Wald distribution with drift rate $\gamma_i$, response threshold $\alpha_i$, and shift $\theta_i$. The collection of parameters $\gamma_i$, $\alpha_i$, and $\theta_i$ for $i=1,\dots,N$ were each randomly drawn from normal distributions with mean and variance set to match the shifted-Wald parameter estimates given in \citet{faulkenberry2018wald}. That is:%
\begin{align*}
  \gamma_i \sim \mathcal{N}(3.91, 0.70^2)\\
  \alpha_i \sim \mathcal{N}(0.92, 0.17^2)\\
  \theta_i \sim \mathcal{N}(0.32, 0.05^2) \; . 
\end{align*}%
To produce the $k$ incongruent trials for each subject $i$, I followed the same sampling scheme, but additionally instantiated a congruity effect $\delta_i$ on the shift parameter $\theta_i$ for each subject $i=1,\dots,N$. Thus the observed incongruent trial RTs for subject $i$ were generated from a shifted-Wald distribution with drift rate $\gamma_i$, response threshold $\alpha_i$, and shift $\theta_i + \delta_i$. Individual differences in the congruity effects were generated following the method of \citet{rouderKumar}, who assumed that the subject-level effects $\delta_i$ were drawn in a hierarchical fashion as $\delta_i \sim \mathcal{N}(\mu_{\delta}, \sigma_{\delta}^2)$, where we must further specify our assumptions on $\mu_{\delta}$ and $\sigma^2_{\delta}$. It is with these assumptions that we can specify the two models which constrain the individual effects $\delta_i$. For data generated under the unconstrained model $\mathcal{M}_u$, the overall mean effect $\mu_{\delta}$ was drawn as $\mu_{\delta} \sim \mathcal{N}(0.1, 0.1^2)$. For data generated under the positive-effects model $\mathcal{M}_+$, $\mu_{\delta}$ was assumed to be drawn from a truncated normal: $\mu_{\delta} \sim \mathcal{N}_+(0.1, 0.05^2)$. For both models, we assumed $\sigma^2_{\delta} \sim \text{Inv-Gamma}(2, 0.03^2)$.

On each simulation run, the collection of $2\cdot k\cdot N$ RTs were submitted to each of the modeling workflows described earlier in the paper. In the first workflow, I applied the default method of \citet{haaf2017}, which assumes the observed RTs are generated via a normal distribution. In the second workflow, I transformed the observed RTs by first subtracting 0.95 times the minimum observed RT (i.e., shifting the distribution to remove the leading edge) and then taking the natural logarithm of the shifted RTs. In all, I completed 200 simulation runs in each of 6 conditions created by systematically varying $N$ and $k$ to reflect common experimental designs in the cognitive and behavioral sciences. Specifically, I crossed $N=20,80$ with $k=50, 100, 200$.

First, let us consider the accuracy of the default and log-transform methods. Tables \ref{tab:accP} and \ref{tab:accU} depict the accuracies obtained from the two methods for each of our 6 experimental conditions for data generated under the positive-effects model $\mathcal{M}_+$ and the unconstrained model $\mathcal{M}_u$, respectively. For data generated under $\mathcal{M}_+$, inference was very accurate for conditions with $N=20$ subjects. Even with smaller numbers of trials (e.g., $k=50$), both the default and log-transform methods were largely correct in their model choices, and there was no obvious difference between the two methods. For larger numbers of subjects ($N=80$), the results were curious. The log-transform method was consistently more accurate than the default method of \citet{haaf2017}, with an accuracy advantage between 5\% and 9\%. Both methods exhibited smallest accuracy when $k=50$, but these accuracies increased with increasing $k$.  For data generated under $\mathcal{M}_u$, performance was consistently at ceiling. Curiously, the worst performance (accuracy = 97\%) occurred in the condition with a small number of participants ($N=20$) each contributing a large number of trials ($k=200$). 

\begin{table}
  \centering \small
  \begin{tabular}{ccccccc}
    \hline
    & & \multicolumn{2}{c}{$N=20$} & & \multicolumn{2}{c}{$N=80$}\\
    & & Default & Log-transform & & Default & Log-transform\\
    \hline
    $k=50$ & &  0.93 & 0.93 & & 0.51 & 0.56\\
    $k=100$ & & 0.92 & 0.93 & & 0.52 & 0.61\\
    $k=200$ & & 0.95 & 0.97 & & 0.71 & 0.79\\
    \hline
    
  \end{tabular}
  \caption{Model choice accuracy for the default \citet{haaf2017} method and the log-transform method, calculated as the proportion of datasets simulated under the positive-effects model $\mathcal{M}_+$ for which $\text{BF}_{+u}>1$.}
  \label{tab:accP}
\end{table}

\begin{table}
  \centering \small
  \begin{tabular}{ccccccc}
    \hline
    & & \multicolumn{2}{c}{$N=20$} & & \multicolumn{2}{c}{$N=80$}\\
    & & Default & Log-transform & & Default & Log-transform\\
    \hline
    $k=50$ & &  0.99 & 0.99 & & 0.99 & 0.99\\
    $k=100$ & & 0.99 & 0.99 & & 0.99 & 0.99\\
    $k=200$ & & 0.97 & 0.97 & & 0.99 & 0.99\\
    \hline
    
  \end{tabular}
  \caption{Model choice accuracy for the default \citet{haaf2017} method and the log-transform method, calculated as the proportion of datasets simulated under the unconstrained model $\mathcal{M}_u$ for which $\text{BF}_{u+}>1$.}
  \label{tab:accU}
\end{table}

While accuracy of the methods is important to assess, the critical claim of this paper is that the default workflow of \citet{haaf2017} produces the same inference as the log-transform method. We can assess this claim empirically by considering the consistency of the inferences obtained by both methods in our simulated datasets. Table \ref{tab:con} shows that these methods exhibit a great deal of consistency. For data generated under $\mathcal{M}_+$, model choice consistency was extremely high (at least 95\%) for conditions with $N=20$. Similar to the results with accuracy above, model choice consistency was not quite as high for conditions with $N=80$, though consistency did increase with increasing $k$. For data generated under $\mathcal{M}_u$, model choice consistency was at ceiling for all conditions.

\begin{table}
  \centering \small
  \begin{tabular}{ccccccc}
    \hline
    & & \multicolumn{2}{c}{Positive-effects} & & \multicolumn{2}{c}{Unconstrained}\\
    & & $N=20$ & $N=80$ & & $N=20$ & $N=80$\\
    \hline
    $k=50$ & &  0.95 & 0.83 & & 0.99 & 0.99\\
    $k=100$ & & 0.95 & 0.85 & & 0.99 & 0.99\\
    $k=200$ & & 0.98 & 0.90 & & 0.99 & 0.99\\
    \hline
    
  \end{tabular}
  \caption{Model choice consistency for the default \citet{haaf2017} method and the log-transform method, calculated as the proportion of datasets for which both methods choose the same model.}
  \label{tab:con}
\end{table}

\section{Conclusion}

The main aim of this paper was to compare the inferences from two methods for assessing the structure of individual differences in behavioral tasks. The first of these two methods was the default Haaf and Rouder (2017) method, which assumes that the observed response times are drawn from a normal distribution. The second was a modified approach where the observed response times are assumed to follow a lognormal distribution. Two case studies and a simulation lead to a common conclusion. Even though observed response times typically exhibit positive skew, the inference we obtain from applying the default \citet{haaf2017} method is practically equivalent to those obtained when we apply a shifted lognormal model to the response times. As we saw in both case studies, applying a shift and then taking the natural logarithm of the observed response times does indeed transform the distribution of observed data into one which is approximately normal. Certainly, the \citet{haaf2017} method works well for this transformed data, but the penalty is in the interpretation. When the observed data is transformed to the log scale, the ``effects'' we see in the data (i.e., differences between the observed data that occur as a function of the experimental manipulation) are now differences in the log scale. Differences in the log scale become multiplicative differences (i.e., quotients) when we transform back to the original scale of the response times. While multiplicative effects can make sense in many contexts, such effects are not typical in the context of effects on response time. Indeed, most typical response time models assume that total response time is the \emph{sum} of its constituent subprocesses \citep{schwarz2001,ashby1980}. As such, it is not clear how one of these behavioral or cognitive effects could reasonably interpreted in a multiplicative context.

Given that (1) the pattern of inference does not change, and (2) the interpretation of estimated effects becomes less clear, there is no compelling reason to reject the normal assumption on response times when applying the \citet{haaf2017} method for investigating individual difference structures in behavioral tasks.

\bibliography{references}
\bibliographystyle{apalike}

\end{document}